\documentclass[10pt]{article}
\usepackage{amsmath,amssymb,tikz}
\usepackage{graphicx}
\usepackage{braket}
\usepackage{bm}

 \allowdisplaybreaks
 \newbox\pippobox

\def\({\left(} \def\){\right)}
\def\[{\left[} \def\]{\right]}

\newcommand{\be}{\begin{equation}}
\newcommand{\ee}{\end{equation}}
\newcommand{\bea}{\begin{eqnarray}}
\newcommand{\eea}{\end{eqnarray}}
\newcommand{\ba}{\begin{eqnarray}}
\newcommand{\ea}{\end{eqnarray}}

\newcommand{\beq}{\begin{equation}}
\newcommand{\eeq}{\end{equation}}
\newcommand{\beqa}{\begin{eqnarray}}
\newcommand{\eeqa}{\end{eqnarray}}
\newcommand{\beqar}{\begin{eqnarray*}}
\newcommand{\eeqar}{\end{eqnarray*}}

\long\def\symbolfootnote[#1]#2{\begingroup%
\def\thefootnote{\fnsymbol{footnote}}\footnote[#1]{#2}\endgroup}

\newcommand{\itp}{\it
Institute of Theoretical Physics, Chinese Academy of Sciences, Beijing 100190}

\newcommand{\yitp}{\it Yukawa Institute for Theoretical Physics (YITP),\\
Kyoto University, Kyoto 606-8502, Japan}
\begin{document}
\thispagestyle{empty}
\begin{center}

~\vspace{20pt}

{\Large\bf Coherent state, local excitation in 2D conformal field theory}

\vspace{25pt}
Wu-Zhong Guo\symbolfootnote[1]{Email:~\sf wuzhong@itp.ac.cn}

\vspace{10pt}\itp

\vspace{10pt}\yitp

\vspace{2cm}

\begin{abstract}
In this paper we discuss the topics concerning the local excitation and coherent state in 2D CFT.
It is shown that the local excitation of primary operator can be taken as a coherent state of the global
conformal group. We also discuss the entanglement property of such state. For rational CFT the entanglement
entropy between the holomorphic and anti-holomorphic sector of the local excitation of some primary operator
 is related to the  quantum dimension of the operator, consistent with previous approach, but by a different
 method. We comment on the possible application of so-defined group coherent state in the holographic view.
\\
We also study the coherent state in the free massless boson field, their time evolution and entanglement property.
We introduce the deformed local excitation and the entangled state constructed by them. It is shown the violation of
Bell inequality for such entangled state.
\end{abstract}

\end{center}

 \newpage

\tableofcontents

\section{Introduction}
The coherent state of radiation field is an important concept in quantum optics after the work of Glauber\cite{Glauber1}\cite{Glauber2}. Glauber constructed the the term coherent state as the the eigenstates of the annihilation operator of the electromagnetic field. This is same as
Schr\"{o}dinger's original idea of the most classical state for quantum harmonic oscillators. It is soon realized that the concept of the
coherent state can be generalized to arbitrary Lie groups\cite{Perelomov}\cite{Gilmore}. The so-called group coherent states have various properties similar to the coherent state of quantum harmonic oscillator. They can also be realized in some real physical systems. Nowadays the coherent states have many different
application, such as condensed matter physics, nuclear physics, thermodynamics and so on, see \cite{Zhang}\cite{Perelomov2}\cite{Klauder}. \\
On the other hand, recently, the locally excited state of a primary operator in field theory and its entanglement property are well studied by many authors, such as in the
free theory\cite{Nozaki:2014hna}\cite{Nozaki:2014uaa}\cite{Nozaki:2015mca}, in 2D conformal field theory with or without boundary\cite{He:2014mwa}\cite{Guo:2015uwa}, in theory with holographic dual\cite{Caputa:2014vaa}\cite{Caputa:2015waa}, and in thermal field\cite{Caputa:2014eta}\cite{Guo:2015uwa}. Even the descendent operator in 2D CFT is also studied by \cite{Chen:2015usa}\cite{Caputa:2015tua}.  The entanglement property means the entanglement between the left and right side of the point where inserting the operator, see the cited papers for more details.  The entanglement  can be explained by
the quasi-particle excitation in the corresponding theory. In this paper we mainly focus on the 2D conformal field theory. In the free theory we study the time evolution of  locally excited state,  which is defined as the ``Glauber''
coherent state, even under a time dependent Hamiltonian.  To discuss the entanglement properties more clear we introduce the so-called deformed local excitation, which can be taken as
exciting the vacuum in a finite region, not a point. For the general conformal field theory any primary operator excitation can be mapped to
a coherent state of the global conformal group $SL(2,\mathcal{C})$. The entanglement entropy between the holomorphic and antiholomorphic sector is
shown to be independent with the coordinate. We also argue the entanglement entropy is related to the quantum dimension of
the primary operator, which is consistent with the approach by replica trick in \cite{He:2014mwa}. The fusion relation of the operators in CFT makes
the corresponding states have the similar property.  The local excitation behaves as the anyons\cite{Wilczek:1982wy}, to obtain the result we use the property of anyons. \\
In section \ref{Coherentstate} we review the definition of coherent state of quantum harmonic oscillator. In section \ref{setup} we set up the
locally excited state in field theory. In section \ref{CoherentFreeTheory} we study the ``Glauber'' coherent state defined in 2D free massless
 boson theory, its physical property and evolution under the free or time-dependent Hamiltonian. In section \ref{Freeentangledcoherentstate}
 the entangled coherent state is discussed in detail. We also define the deformed coherent state, and the entangled states constructed by it.
 The violation of Bell inequality for the entangled state is also briefly discussed. In section \ref{GCFT} the relation between a group coherent state
and local excitation is given for general conformal field theory. In section \ref{EEForRCFT} the entanglement entropy of the local excitation is obtained by considering the fusion relation of primary operator. Section  \ref{Comment} gives short comment on the holographic view on the coherent state. Last section is the conclusion and further application of this paper.

\section{Coherent state}\label{Coherentstate}

The concept of what is now called coherent state was first introduced by Schr\"{o}dinger for quantum harmonic oscillator. The quantum harmonic oscillator
can be constructed by the annihilation operator $a$ and creation operator $a^{\dag}$, which satisfies the commutation relation $[a,a^\dag]=I$. The set of
operator $\{a,a^\dag, \hat n\equiv a^\dag a, I\}$ spans a Lie algebra, which are generators of Heisenberg-Weyl group $H_4$. The Hilbert space is spanned by the eigenstates of the number operator $\hat n$, and the vacuum is defined by $\hat n \ket{0}=0$. There are three different ways to define the coherent state for quantum harmonic oscillator\cite{Glauber1}. Here we follow \cite{Perelomov}\cite{Zhang} to construct
the coherent states, which is easy to generalize to any Lie group.\\
We choose the vacuum $\ket{0}$ as the reference state. A general element of the Lie algebra can be written as
\begin{eqnarray}
\theta I+ \xi \hat n+\alpha a+ \alpha^* a^\dag,
\end{eqnarray}
where $\theta, \xi, \alpha$ are arbitrary complex number. For $H_4$ there is a subgroup which leaves the reference state $\ket{0}$ invariant up to some phase. This subgroup $G$ is generated by $\{\hat n,I\}$. The
corresponding group element is obtained by the exponential mapping
\begin{eqnarray}
h=e^{i\theta I+ i\xi \hat n}.
\end{eqnarray}
Thus we have
\begin{eqnarray}
h\ket{0}=e^{i\theta}\ket{0}.
\end{eqnarray}
The coherent states are related to the coset space with respect to subgroup $G_0$, i.e., $H_4/G_0$. The elements in such coset group are
\begin{eqnarray}
D(\alpha)=e^{\alpha a^\dag-a^* a}.
\end{eqnarray}
Thus the coherent states are the set of vectors $\ket{\alpha}$ with
\begin{eqnarray}
\ket{\alpha}=D(\alpha)\ket{0}.
\end{eqnarray}
This is one of definitions of the coherent state given by Glauber \cite{Glauber1}. It is well known that such state are nonorthogonal with each other,
\begin{eqnarray}
\langle \alpha| \alpha'\rangle=e^{\alpha^*\alpha'-\frac{1}{2}(\alpha^*\alpha+\alpha'^*\alpha')}.
\end{eqnarray}
It is also an eigenstates of the annihilation operator $a$,
\begin{eqnarray}
a\ket{\alpha}=\alpha \ket{\alpha}.
\end{eqnarray}

\section{Set-up of local excitation}\label{setup}
In this section we will study the locally excited state of some primary operator $O$ in 2D CFT. We assume the Hamitonian of the CFT is $H_0$. One could obtain the local excitation by interaction between specific operator $O$ and external time-dependent source, via the Hamiltonian $H_{int}$,
\begin{eqnarray}
H_{int}= \epsilon(t) O(x),
\end{eqnarray}
where the operator is inserted at $x$. We will assume that the coupling function $\epsilon(t)$ is small, the external source is turned on during time $0<t<T$. With this one could get the time evolution operator $U$ in the interaction representation as
\begin{eqnarray}
U(t,0)&=&e^{-i H^I_{int}}\nonumber \\
&=& e^{-i \int_0^{t} \epsilon(t) O^I(x,t)dt},
\end{eqnarray}
where the superscript ``I'' means the corresponding operator in the interaction representation.
The initial state is assumed as the vacuum $\ket{0}$. So final state after the interaction is
\begin{eqnarray}
\ket{T}&=&U(T,0)\ket{0}\\ \nonumber
&&\simeq \ket{0}-i\int_0^{T}\epsilon(t) O^I(x,t)dt \ket{0}\\ \nonumber
&&= \ket{0} +N O(x)\ket{0},
\end{eqnarray}
where $N$ is constant. This state can be seen as a perturbation of the vacuum. In the following we will focus on the perturbation part and  consider the
time evolution of the state $O(x)\ket{0}$ by $U_0=e^{-iH_0 t}$. The state at time $t$ is
\begin{eqnarray}
\ket{\psi(x,t)}=U_0 O(x)\ket{0}= O(x,t)\ket{0},
\end{eqnarray}
where $O(x,t)\equiv U_0 O(x)U^{\dag}_0$. If we continue to Euclidean space, with $t\to -i\tau$, the space can be described by a complex coordinate $\omega=\tau+i x$. The operator $O(x,t)$ is denoted as $O(\omega,\bar \omega)$.\\
\section{Free boson}\label{CoherentFreeTheory}
The free boson is the simplest example of 2D CFT, which is also the building blocks of more complicated models. Let $\phi(x,t)$ be a free Bose field defined on cylinder of circumference $L$, with the condition $\phi(x+L,t)=\phi(x,t)$. The action is
\begin{eqnarray}\label{freetheorymassless}
S=\frac{1}{8\pi}\int dxdt \partial_\mu \phi \partial^\mu \phi.
\end{eqnarray}
The Hamiltonian is
\begin{eqnarray}\label{HforFreescalar}
H=\frac{1}{8\pi}\int dx [(\partial_t \phi)^2+(\partial_x \phi)^2].
\end{eqnarray}
By the canonical quantization the mode expansions of $\phi$ at time slice $t=0$ is
\begin{eqnarray}\label{Freebosonmode}
\phi(x,t=0)=\phi_0+i\sum_{n\ne 0}\frac{1}{n}(a_n-\bar a_{-n})e^{2\pi inx/L},
\end{eqnarray}
with the associated commutation relations
\begin{eqnarray}
[a_n,a_m]=n \delta_{n+m},\ \ \ [a_n,\bar a_m]=0,\ \ \ [\bar a_n,\bar a_m]=n\delta_{n+m}.
\end{eqnarray}
Then the Hamiltonian (\ref{HforFreescalar}) is
\begin{eqnarray}
H=\frac{2\pi}{L}\pi_0^2+\frac{2\pi}{L}\sum_{n\ne 0}(a_{-n}a_n+\bar a_{-n}\bar a_n),
\end{eqnarray}
where $\pi_0$ is the momentum conjugated to $\phi_0$, which satisfies $[\pi_0,\phi_0]=i$. The time evolution of $\phi$ is
\begin{eqnarray}\label{Freebosonmodeattime}
\phi(x,t)=A+\bar A,
\end{eqnarray}
where
\begin{eqnarray}\label{aandabar}
&&A=\phi_0+\frac{4\pi \pi_0 t}{L}+i\sum_{n\ne 0}\frac{1}{n}(a_n e^{2\pi i n(x-t)/L}),\\
&&\bar A=-i\sum_{n\ne 0}\frac{1}{n}(\bar a_{-n} e^{2\pi i n(x+t)/L})
\end{eqnarray}

\subsection{Coherent state for Boson field}
The free boson field can be seen as infinite decoupled quantum harmonic oscillators.  So we could define the coherent state by
\begin{eqnarray}\label{CoherentFreeBoson}
\ket{\alpha}=e^{i\phi(x,t=0)}\ket{0}=N_{\alpha} e^{\sum_{n>0}\frac{1}{n}(a_{-n}e^{-2\pi i nx/L}+\bar a_{-n}e^{2\pi i nx/L})}\ket{0},
\end{eqnarray}
where $\alpha$ refers to the parameters associated to $a_n$ and $\bar a_n$ in (\ref{Freebosonmode}),
\begin{eqnarray}
N_\alpha
\equiv e^{-\sum_{n>0}\frac{1}{n}}.
\end{eqnarray}
It is also a local excitation by the operator $O=e^{i\phi}$. If the Hamiltonian is the free one (\ref{HforFreescalar}), at any
time $t$, the state (\ref{CoherentFreeBoson}) evolutes to
\begin{eqnarray}
\ket{\alpha(t)}=e^{i\phi(x,t)}\ket{0}=N_\alpha e^{\sum_{n>0}\frac{1}{n}(a_{-n}e^{-2\pi i n(x-t)/L}+\bar a_{-n}e^{2\pi i n(x+t)/L})}\ket{0},
\end{eqnarray}
where $\alpha(t)$ denotes the parameters associated to $a_n$ and $\bar a_n$ . Thus once we obtain a  coherent state like (\ref{CoherentFreeBoson}), it will always be a coherent state. In fact we could abandon the condition that the state evolutes by the free Hamiltonian (\ref{HforFreescalar}), but use the more general time-dependent Hamiltonian,
\begin{eqnarray}\label{firstorderH}
H_1&=&\omega(t) H+\sum_{n} \lambda_n(t)\int dx \phi(x,t=0)e^{2\pi in x/L}+f(t)\nonumber \\
&=&\frac{1}{8\pi} \int dx [\omega(t)((\partial_t \phi)^2+(\partial_x \phi)^2)+8\pi \Lambda(x,t)\phi(x,t=0)]+f(t),
\end{eqnarray}
where $\omega(t)$, $\lambda_n(t)$, and $f(t)$ are time-dependent function, we also define $\Lambda(x,t)$
\begin{eqnarray}
\Lambda(x,t)=\sum_n \lambda_n(t)e^{2i\pi nx/L},
\end{eqnarray}
as the inverse Fourier transformation of $\lambda_n$.\\
Let's work in the Heisenberg picture, and assume the initial state is the coherent state (\ref{CoherentFreeBoson}). The time evolution of the operator $a_n(t)$  satisfy
\begin{eqnarray}\label{HeisenbergEOM}
i\frac{da_n(t)}{dt}=[a_n(t),\hat H_1],
\end{eqnarray}
where $\hat H_1$ is the corresponding operator of the Hamiltonian (\ref{firstorderH}), and $a_n(t=0)=a_n$.
We want that the state (\ref{CoherentFreeBoson}) at time $t$ is still a coherent state, which means it is still eigenstate of the operator $a_n(t)$.
Taking (\ref{firstorderH}) into the Heisenberg equation (\ref{HeisenbergEOM}) we have
\begin{eqnarray}
\frac{da_n(t)}{dt}=-i\frac{2n\pi\omega(t)}{L}a_n(t)+L\lambda_{-n}(t).
\end{eqnarray}
One could obtain the solution as
\begin{eqnarray}
a_n(t)=e^{-i\psi_n(t)}a_n+\alpha_n(t),
\end{eqnarray}
where
\begin{eqnarray}
\psi_n(t)=\frac{2\pi n}{L}\int_0^t \omega(t')dt',\ \ \alpha_n(t)=Le^{-i\psi_n(t)} \int_0^tdt' \lambda_{-n}(t')e^{i\psi_n(t)}.
\end{eqnarray}
We could also obtain the equation of motion for $\bar a_n$,
\begin{eqnarray}
\frac{d\bar a_n(t)}{dt}=-i\frac{2n\pi\omega(t)}{L}\bar a_n(t)+L\lambda_{n}(t),
\end{eqnarray}
the solution
\begin{eqnarray}
\bar a_n=e^{-i\psi_n(t)}a_n+\bar \alpha_n(t),
\end{eqnarray}
with
\begin{eqnarray}
 \bar \alpha_n(t)=Le^{-i\psi_n(t)} \int_0^tdt' \lambda_{n}(t')e^{i\psi_n(t)}.
\end{eqnarray}
As $a_n(t)$ is linear function of $a_n$ (\ref{CoherentFreeBoson}) is still an eigenstate of $a_{n}(t)$£¬ with eigenvalue $\beta_n(t)$
\begin{eqnarray}
\beta_{+n}(t)=e^{-2\pi i nx/L-i\psi_n}+\alpha_n(t).
\end{eqnarray}
So is $\bar a_n(t)$, with the eigenvalue
\begin{eqnarray}
\bar \beta_{+n}=e^{2\pi i nx/L-i\psi_n}+\bar \alpha_n(t).
\end{eqnarray}
For the mode $\phi_0$, the result is
\begin{eqnarray}
\phi_0(t)=\phi_0+\frac{4\pi \pi_0}{L}\int_0^t\omega(t) dt.
\end{eqnarray}
 In the Schr\"{o}dinger picture it is not easy to write down the final state after the evolution. But since it should be the eigenstate of $a_n$, one could guess it must have the form
 \begin{eqnarray}
 \ket{\alpha(t)}= e^{i\Phi_+(t)}N_{+\beta} e^{\sum_{n>0} \frac{1}{n}(a_{-n}\beta_{+n}(t)+\bar a_{-n}\bar \beta_{+n}(t)}\ket{0},
\end{eqnarray}
where $N_{+\beta}$ is the normalization constant, $\Phi_{+}(t)$ is a real function.\\
To be useful later we also define the coherent state
\begin{eqnarray}
\ket{-\alpha}=e^{-i\phi(x,t=0)}\ket{0},
\end{eqnarray}
which is different from (\ref{CoherentFreeBoson}) just by the sign. Following the same method we used above one could get the final state under the
evolution of the time dependent Hamiltonian (\ref{firstorderH}),
\begin{eqnarray}
\ket{\alpha(t)}=e^{i\Phi_-(t)}N_{-\beta} e^{\sum_{n>0} \frac{1}{n}(a_{-n}\beta_{-n}(t)+\bar a_{-n}\bar \beta_{-n}(t)}\ket{0},
\end{eqnarray}
where we also define
\begin{eqnarray}
\beta_{-n}(t)=-e^{-2\pi i nx/L-i\psi_n}+\alpha_n(t), \bar \beta_{-n}=-e^{2\pi i nx/L-i\psi_n}+\bar \alpha_n(t),
\end{eqnarray}
and $N_{-\beta}$ is the normalization constant.
\subsection{Physical behavior of such state}\label{SecClassical}
The coherent state defined in the quantum harmonic oscillator has the good property to describe its classical motion. This can be seen by calculating
the expectation value of the position and momentum operators in the coherent state. They indeed behave as the position and momentum of the classical
harmonic oscillator\cite{Zhang}. This property is further generalized to more complex case in the classical limit $\hbar\to 0$\cite{Hepp} or large $N$ limit\cite{Yaffe}. Here we would like to show
that the coherent state (\ref{CoherentFreeBoson}) behaves as qusi-particles, which is consistent with the physical intuition for the local excitation.\\
Unlike the harmonic oscillator the coordinates of the field theory  phase space are usually not good physical observables. To see the ``classical'' behavior of the coherent state (\ref{CoherentFreeBoson}), we would like to use the energy and momentum density $\mathcal{H}$,$\mathcal{P}$  as the observables. We will work in the Schr\"odinger picture. In the free field theory they are defined by
\begin{eqnarray}\label{Hdensity}
\mathcal{H}(x)&=&\frac{1}{8\pi}[(\partial_t \phi(x,t=0))^2+(\partial_x\phi(x,t=0))^2]\nonumber \\
&=&\frac{\pi}{2L^2}\sum_{n,m}[a_na_me^{2i\pi x(m+n)/L }+\bar a_n \bar a_me^{-2i\pi x(m+n)/L }]
\end{eqnarray}
\begin{eqnarray}\label{Pdensity}
\mathcal{P}(x)&=&-\frac{1}{8\pi}[\partial_t \phi(x,t=0)\partial_x\phi(x,t=0)]\nonumber \\
&=&\frac{\pi}{2L^2}\sum_{n,m}[a_na_me^{2i\pi x(m+n)/L }-\bar a_n \bar a_me^{-2i\pi x(m+n)/L }].
\end{eqnarray}
At time t the coherent state evolutes to (\ref{Freebosonmodeattime}). We assume it is properly normalized. It is eigenstate of $a_n$ ($n > 0$),\\
\begin{eqnarray}
a_n \ket{\alpha(t)}=- e^{-2\pi in(x-t)/L}\ket{\alpha(t)},
\end{eqnarray}
also
\begin{eqnarray}
\bra{\alpha^*(t)}a_{-n}=-e^{2\pi i n(x-t)/L}.
\end{eqnarray}
Let's rewrite (\ref{Hdensity}) as
\begin{eqnarray}
\mathcal{H}(x')&=&\frac{\pi}{2L^2}\sum_{k \ge 0,l\ge 0}\Big[a_ka_l e^{2\pi ix'(k+l)/L}+a_{-k}a_{-l} e^{-2\pi ix'(k+l)/L}+a_{-k}a_l e^{2\pi ix'(-k+l)/L}+a_{-l}a_k e^{2\pi ix'(k+l)/L}\Big]\nonumber \\
&+&\frac{\pi}{2L^2}\sum_{k \ge 0,l\ge 0}\Big[\bar a_k \bar a_l e^{2\pi ix'(k+l)/L}+\bar a_{-k}\bar a_{-l} e^{-2\pi ix'(k+l)/L}+\bar a_{-k}\bar a_l e^{2\pi ix'(-k+l)/L}+\bar a_{-l}\bar a_k e^{2\pi ix'(k+l)/L}\Big]\nonumber \\
&+&C,
\end{eqnarray}
where  $C$ is as constant. So we obtain
\begin{eqnarray}\label{energydenstiyvalue}
\bra{\alpha^*(t)}\mathcal{H}(x')\ket{\alpha(t)}&=&\frac{2\pi}{L^2}\sum_{k \ge 0,l\ge 0}[\sin(\frac{2\pi k(x'-x+t)}{L})\sin(\frac{2\pi l(x'-x+t)}{L})]\\ \nonumber
&=&\frac{1}{8\pi}[\delta(x'-x+t)^2+\delta(x'-x-t)^2].
\end{eqnarray}
The energy density is located at the position $x'=x-t$ and $x'=x+t$ as we can see from the delta function in the result. At $t=0$ the energy density is produced at x, where the local operator is inserted. At time $t$ the energy density propagates to $x'=x+t$ and $x'=x-t$. Thus  the velocity of the propagation of energy is $v=1$.\\
For $\mathcal{P}$ it is straightforward to get
\begin{eqnarray}\label{momentumdensityvalue}
\bra{\alpha(t)^*}\mathcal{P}(x')\ket{\alpha(t)}=\frac{1}{8\pi} [\delta(x'-x+t)^2-\delta(x'-x-t)^2],
\end{eqnarray}
the only difference with $\langle\mathcal{H}\rangle$ is the minus sign, which means that the direction of the energy propagation at $x'=x-t$ and
$x'=x+t$ is converse. This is consistent with our physical intuition on the local excitation.

\section{Entangled coherent state}\label{Freeentangledcoherentstate}
As we show in the above section the coherent state (\ref{Freebosonmodeattime}) can be explained as two quasi-particles which spread to two different direction. But these two quasi-particle are not entangled with each other, since the wave function can be written as a product state. To obtain a
entangled state one could use the operator $O=\lambda_1 e^{i\phi}+\lambda_2 e^{-i\phi}$. Here we will focus on the typical one $O=e^{i\phi}\pm e^{-i\phi}$. Thus at $t=0$ we get a state
\begin{eqnarray}\label{entangledscalartimeZero}
\ket{\alpha,-\alpha}= [e^{i\phi(x,t=0)}\pm e^{-i\phi(x,t=0)}]\ket{0}.
\end{eqnarray}
At time $t$ it becomes the state
\begin{eqnarray}\label{Bellcoherentstatetime}
\ket{\alpha(t),-\alpha(t)}= [e^{i\phi(x,t)}\pm e^{-i\phi(x,t)}]\ket{0},
\end{eqnarray}
if we assume the Hamiltonian is the  free one (\ref{HforFreescalar}).
One can also show that the energy and momentum density expectation value is same as (\ref{energydenstiyvalue}) and (\ref{momentumdensityvalue}), which means that we still obtain two quasi-particles. This can be seen as the entangled coherent state, which is studied by many authors, see the review \cite{Sanders} and the references therein. These state are also called by quasi-Bell state, which has two nonorthogonal states $\ket{\alpha_1}$
and  $\ket{\alpha_2}$ such that $\langle \alpha_1|\alpha_2\rangle =\kappa$. We have the following quasi-Bell states
\begin{eqnarray}\label{Coherententangledgenral}
\ket{\Psi_{\pm}}=N_{\pm} (\ket{\alpha_1}_A\ket{\alpha_1}_B\pm \ket{\alpha_2}_A\ket{\alpha_2}_B),
\end{eqnarray}
where $N_{\pm}$ is the normalization constant, $A,B$ are two subsystem. The entanglement entropy of such states are
\begin{eqnarray}
&&S(\ket{\Psi_+})=-\frac{(1+\kappa)^2}{2(1+\kappa^2)}\log \frac{(1+\kappa)^2}{2(1+\kappa^2)}-\frac{(1-\kappa)^2}{2(1+\kappa^2)}\log \frac{(1-\kappa)^2}{2(1+\kappa^2)}, \\
&&S(\ket{\Psi_-})=\log 2.
\end{eqnarray}
Thus $\ket{\Psi_-}$ is always a maximal entangled state, however $\ket{\Psi_+}$ is maximal entangled states only for $\kappa=0$ or $\infty$.\\
The entangled coherent state (\ref{Bellcoherentstatetime}) can be rewritten as
\begin{eqnarray}\label{Entangledcoherentscalr}
\ket{\alpha(t),-\alpha(t)}=\ket{\alpha}\ket{\bar \alpha}\pm \ket{-\alpha}|-\bar \alpha\rangle,
\end{eqnarray}
where we define $\ket{\pm\alpha}$ and $\ket{\pm\bar \alpha}$ as
\begin{eqnarray}\label{stateleftandright}
\ket{\alpha}= e^{i A}\ket{0}, \ket{\bar \alpha}=e^{i \bar A}\ket{0}, \ket{-\alpha}= e^{-i A}\ket{0}, \ket{-\bar \alpha}=e^{-i \bar A}\ket{0}.
\end{eqnarray}
We refer the states $\ket{\pm \alpha}$ and $\ket{\pm \bar \alpha}$ to the wave function of quasi-particles located at $x'=x-t$ and $x'=x+t$ respectively.
They are orthogonal to each other, i.e., $\bra{\alpha^*}|\bar \alpha\rangle=0$. Such states are properly normalized
and
\begin{eqnarray}\label{kappa}
\kappa_\alpha\equiv\langle \alpha^*\ket{-\alpha}=\langle \bar\alpha^*\ket{-\bar\alpha}=e^{-\sum_{n=1}^{\infty}2\frac{1}{n}}.
\end{eqnarray}
Thus the state (\ref{Entangledcoherentscalr}) is as the same form as (\ref{Coherententangledgenral}), with the normalization constant $N_{\pm}=1/\sqrt{2\pm2 k_\alpha^2}$. Since $\kappa_\alpha \to 0$ we get the entanglement entropy is $\log 2$, which is maximal entangled state.\\
One could also assume the initial state (\ref{entangledscalartimeZero}) evolves under the time-dependent Hamiltonian (\ref{firstorderH}). At time $t$
the entangled state becomes
\begin{eqnarray}\label{estateundertimedepend}
\ket{\tilde\alpha(t),-\tilde\alpha(t)}=e^{i\Phi_+}N_{+\beta}\ket{\tilde{\alpha}}\ket{{\bar{\tilde \alpha}}} \pm e^{i\Phi_-}N_{-\beta}\ket{-\tilde{\alpha}}\ket{{-\bar{\tilde \alpha}}},
\end{eqnarray}
where
\begin{eqnarray}
\ket{ \pm\tilde \alpha}= e^{\sum_{n>0} \frac{1}{n}a_{-n}\beta_{\pm n}(t)}\ket{0},
\ket{ \pm\bar{\tilde \alpha}}= e^{\sum_{n>0} \frac{1}{n}\bar a_{-n}\bar \beta_{\pm n}(t)}\ket{0},
\end{eqnarray}
this is a complex expression which depends on the parameters $\omega(t)$, $\Lambda(x,t)$, and the coordinates. In the case when $\Lambda(x,t)$ is zero, i.e., there is no linear source, the result is same, the state is maximally entangled. Because in this case $\beta_{\pm n}$ are just phase, they satisfy
$\beta_{+n}*\beta_{-n}=1$, thus the product of the state $\bra{\tilde {\alpha}*}$ and $\ket{-\tilde \alpha}$, $\bra{\tilde {\alpha}}-\tilde \alpha\rangle$
is still $\kappa_\alpha\to 0$ (\ref{kappa}). But when $\Lambda(x,t)\ne 0$, the result will change, $\kappa_\alpha \ne 0$. The states $\ket{\Psi_{\pm}}$ are not the standard Bell state. Even for the $-$ sign case of (\ref{estateundertimedepend}) it is not a maximally entangled state. We could say that the linear source would decrease the entanglement in this example.   In some experiment one could construct the
entangled state similar as (\ref{Entangledcoherentscalr}) for photons, it may be  interesting to work out the entanglement response to source perturbation
in such real system.
\subsection{Deformed local excitation}\label{Deformedlocaexcitation}
The local excitation state (\ref{CoherentFreeBoson}) is defined at some spcetime point. But in the physical experiment one do not expect
the measure or operation is located at a point, but some spacetime region. This is also one of the motivation to introduce the observables in finite region $O$ in algebraic QFT\cite{Haag}. The strategy is using the test functions $f(x)$, which is $C^{\infty}$ functions of rapid decrease. And all the observables are
constructed by the smoothed field $\int f(x) \phi(x)dx$. One could define the observables belonging to some subregion $O$ by requiring the support of $f(x)$ is $O$. Thus we define the deformed local excitation located at some region $O$ for free scalar by
\begin{eqnarray}\label{Deformedcoherentstate}
\ket{\alpha}_D=N e^{\int_O f(x) \phi(x)dx}\ket{0},
\end{eqnarray}
where we use the subscript of the integral $O$ to denote that the support of the test function $f(x)$ is $O$, $N$ is the normalization constant. When $f(x)=i\delta(x-x_0)$ we get the local excitation (\ref{CoherentFreeBoson}). Here we assume the region $O$ is a interval whose middle point is $x_0$, and $f(x)$ is real. By using the mode expansion (\ref{Freebosonmode}) we could rewrite (\ref{Deformedcoherentstate}) as
\begin{eqnarray}
\ket{\alpha}_D=N e^{L\phi_0 f_0 +iL\sum_{n\neq 0}\frac{1}{n}f_{-n}(a_n-\bar a_{-n})},
\end{eqnarray}
where $f_n$ is the Fourier transformation of the test function $f(x)$,
\begin{eqnarray}\label{Fouriertestfunction}
f_n=\frac{1}{L} \int_O dx f(x)e^{-2i\pi nx/L}.
\end{eqnarray}
If we assume the Hamiltonian is (\ref{HforFreescalar}), at time $t$ this state becomes
\begin{eqnarray}
\ket{\alpha(t)}_D=N(t)e^{Lf_0(\phi_0+\frac{4\pi \pi_0 t}{L})+iL\sum_{n\ne 0}\frac{1}{n}f_{-n} [a_ne^{-2\pi int/L}-\bar a_{-n}e^{2\pi int/L}]},
\end{eqnarray}
where $N(t)$ is the normalization constant. One could calculate the energy and momentum density as section \ref{SecClassical}, the result is
\begin{eqnarray}
_D\bra{\alpha(t)^*}\mathcal{H}(x')\ket{\alpha(t)}_D&{\propto}& \sum_{k,l}f_k f_l e^{2\pi i (k+l)(x'+t)/L}+\sum_{k,l}f_k f_l e^{2\pi i (k+l)(x'-t)/L}.
\end{eqnarray}
By using (\ref{Fouriertestfunction}) one could see $\langle \mathcal{H}(x')\rangle$ is non-vanishing if and only if $x'=x-t$ or $x'=x+t$ for some $x\in O$. The energy
density is proportional to $f(x)^2$. Thus  at time $t$ the energy density is located at region $O+t$ and $O-t$. For the momentum density
 we could get similar result, but the momentum density at $O+t$ and $O-t$ have converse sign. The local excitation behave as
two wave packets which spread in two converse direction. But the density is not divergent. When $f(x)=i\delta(x-x_0)$, we obtain the result(\ref{energydenstiyvalue})(\ref{momentumdensityvalue}).\\
It is straightforward to define the entangled state by the deformed local excitation state(\ref{Deformedcoherentstate}),
\begin{eqnarray}\label{DeformedCoherentEntangled}
\ket{\alpha(t),\pm\alpha(t)}_D=N_{\pm}[e^{\int_O f(x) \phi(x,t)dx}\pm e^{-\int_O f(x) \phi(x,t)dx}]\ket{0},
\end{eqnarray}
where $N_{\pm}$ is the normalization constant. This state also behaves as two wave packets located at region $O+t$ and $O-t$. These two wave packets
are described by
\begin{eqnarray}\label{Deformedentangledcoherentstate}
\ket{\alpha(t),-\alpha(t)}_{D\pm}=N_{\pm}(\ket{\alpha}_D\ket{\bar \alpha}_D\pm \ket{-\alpha}_D\ket{-\bar \alpha}_D),
\end{eqnarray}
with
\begin{eqnarray}
\ket{\pm\alpha}_D= e^{\pm \int_O f(x)A}\ket{0},
\end{eqnarray}
\begin{eqnarray}
\ket{\pm\bar \alpha}_D= e^{\pm \int_O f(x)\bar A}\ket{0},
\end{eqnarray}
where $A$  and $\bar A$ are defined by (\ref{aandabar}).
We have
\begin{eqnarray}
\kappa_D=_D\bra{\alpha^*}-\alpha\rangle_D=_D\bra{-\alpha^*}\alpha\rangle_D= e^{-2L^2 \sum_{n>0} \frac{1}{n} f_nf_n^*}.
\end{eqnarray}
The normalization constant $N_{\pm}=1/\sqrt{2+2\kappa_D^2}$. Thus we could obtain the entanglement entropy for the state $\ket{\alpha(t),-\alpha(t)}_{D-}$ is $\log 2$, which is still maximally entangled state. But the entanglement entropy of  state $\ket{\alpha(t),-\alpha(t)}_{D+}$ is
\begin{eqnarray}
S_{+}=-\frac{(1+\kappa_D)^2}{2(1+\kappa_D^2)}\log \frac{(1+\kappa_D)^2}{2(1+\kappa_D^2)}-\frac{(1-\kappa_D)^2}{2(1+\kappa_D^2)}\log \frac{(1-\kappa_D)^2}{2(1+\kappa_D^2)}.
\end{eqnarray}

\subsection{Violation of Bell's inequality}
The Bell inequality concern measurement made by observers located at two spacelike region, which is used to distinguish the local realism theory and
quantum mechanics. The violation of Bell inequality reflects the entanglement property of the state. Even in the vacuum of a quantum field theory it is
shown by \cite{Summers1}\cite{Summers2} that the Bell inequality is violated. We would like to study the Bell inequality for the locally excited states. For some local excitation, e.g., (\ref{Entangledcoherentscalr}), we expect the violation of Bell inequality, since such states are entangled. We will mainly concern on the deformed local excitation $\ket{\alpha(t),-\alpha(t)}_{D-}$ (\ref{Deformedentangledcoherentstate}) in the free theory. We will follow the construction
of the Bell operator of two-particle nonorthogonal states in \cite{Mann}. The state $\ket{\alpha(t),-\alpha(t)}_{D-}$ can be expressed in the Schmidt form
\begin{eqnarray}
\ket{\alpha(t),-\alpha(t)}_{D-}=\lambda_1 \ket{+}_A\ket{+}_B+\lambda_2 \ket{-}_A \ket{-}_B,
\end{eqnarray}
with $|\lambda_1|^2+|\lambda_2|^2=1$. As we have shown in section (\ref{Deformedlocaexcitation}) the state $\ket{\alpha(t),-\alpha(t)}_{D-}$ is
the ``local'' state defined in the region $O+t$ and $O-t$. If we assume Alice and Bob do the experiment, they should make the measure inside the region
$O+t$ and $O-t$ respectively, or they will get a trivial result. For simplicity we assume Alice and Bob choose the Hermitian operators which are constructed by projectors for the states $\ket{\pm}_{A(B)}$.\\
One could obtain the orthogonal basis for Alice and Bob
\begin{eqnarray}
&&\ket{+}_A=N_{A+}(\ket{\alpha}+\ket{-\alpha}),\quad \ket{-}_{A}=N_{A-}(\ket{\alpha}-\ket{-\alpha}), \nonumber \\
&&\ket{+}_{B}=N_{B+}(\ket{\bar \alpha}-\ket{-\bar \alpha}) \quad , \ket{-}_B=N_{B+}(\ket{\bar \alpha}+\ket{-\bar \alpha}),
\end{eqnarray}
where $N_{A\pm}=1/\sqrt{2\pm2\kappa_D}$, $N_{B\mp}=1/\sqrt{2\pm 2\kappa_D}$, $\lambda_1=\lambda_2=\frac{1}{\sqrt{2}}$. The Hermitian operators which have eigenvalues $\pm 1$ can be written as the general form
\begin{eqnarray}
\hat \Theta=\cos \varphi [\ket{+}\bra{+}-\ket{-}\bra{-}]+\sin \varphi [e^{i\theta}\ket{+}\bra{-}+e^{-i\theta}\ket{-}\bra{+}].
\end{eqnarray}
With this one could get the Bell operator
\begin{eqnarray}
\hat B= \hat \Theta_A \hat \Theta_B+\hat \Theta_A \hat {\Theta '}_B
+\hat \Theta '_A\hat \Theta_B-\hat \Theta '_A\hat \Theta '_B
\end{eqnarray}
For the choices
\begin{eqnarray}
&&\varphi_A=0, \varphi_A'=\pi/2,\nonumber \\
&&\varphi_B=\varphi_B'-\frac{\pi}{2}=\varphi_0, \theta=0
\end{eqnarray}
The expectation value of the Bell operator for the deformed entanglement coherent state $\ket{\alpha(t),-\alpha(t)}_{D-}$ is
\begin{eqnarray}
B\equiv  _{D-}\bra{\alpha(t)^*,-\alpha(t)^*} \hat{B}\ket{\alpha(t),-\alpha(t)}_{D-} =2(\sin \varphi_0+ \cos \varphi_0),
\end{eqnarray}
the maximal value is $2\sqrt{2}$ when $\varphi_0=\pi/4$. The state $\ket{\alpha(t),-\alpha(t)}_{D-}$ violates the Bell's inequality maximally.
One could follow the similar process and obtain the violation of Bell inequality for the state $\ket{\alpha(t),-\alpha(t)}_{D+}$, not maximally of course.\\
It is known that the Bell's inequalities are violated in the vacuum state of quantum field theory for two spacelike regions, see, e.g., \cite{Summers1}\cite{Summers2}. Here we show that the Bell's inequalities are violated for the local excitation state of the special operator.
In the vacuum the maximal expectation value of the Bell operator ia s decreasing function of the spacelike distance between two regions because of
the clustering property. Here Alice and Bob do the experiment at region $O+t$ and $O-t$, which are spacelike. The expectation value of the Bell operator
is not related to the distance. This is also consistent with our conclusion that the entanglement entropy of the local excitation is independent with
the coordinates, but just the information of the operator, see the next section. It can be seen as the topological property of the local excitation.
\section{General CFT}\label{GCFT}
The free theory can be seen as infinite decoupled quantum harmonic oscillator. The coherent state defined by (\ref{CoherentFreeBoson}) and
the entangled coherent state defined by (\ref{Entangledcoherentscalr}) is a
generalization of the one quantum harmonic oscillator. But these so-called vertex operators are just one kind of the primary operators in the
free boson CFT. One would wonder whether the other operators, e.g., $\partial_x \phi$, can be taken as the coherent state. In the interacting field theory we don't have such well defined coherent as the generalization of quantum harmonic oscillator. In
this section we would like the discuss  the local excitation by primary operators in general CFT in two dimension, which can also be taken as
generalized coherent state. To see this we need the radial quantization of CFT.
\subsection{Radial quantization}
This topic has been the standard contents in the textbook. For completeness here we list some contents that we will use in the following.
One theory is initially defined on infinite spacetime cylinder, with $t \in (-\infty,+\infty)$, $x\in (0, L)$. In the Euclidian space this cylinder is
described by the complex coordinate $\omega= \tau+i x$. By the conformal transformation
\begin{eqnarray}\label{radialmap}
z=e^{2\pi \omega/L},
\end{eqnarray}
the cylinder is mapped to a complex plane, and $\tau\to -\infty$ is mapped to  the origin $z=0$, the time slice $\tau=C$($C\ne 0$) is mapped to a circle with radius $r=e^{2\pi \tau/L}$. We will
construct Hilbert space in the complex, and assume there is a vacuum state $\ket{0}$. The Hilbert space is obtained by application of ``creation operators'' on vacuum state. The asymptotic state, which is defined at time $\tau\to \infty $ on the cylinder, now reduced to operator a operator
at the origin acting on the vacuum $\ket{0}$. We assume there are a series of primary operators in this theory. Choosing one of these operator, e.g., $\phi(z,\bar z)$ which has the conformal dimension $(h,\bar h)$, we obtain the state
\begin{eqnarray}\label{highestrepresentation}
\ket{h,\bar h}=\lim_{z,\bar z\to 0}\phi(z,\bar z)\ket{0}.
\end{eqnarray}
For the free theory one could choose the operators $\mathcal{V}_\alpha(z,\bar z)= e^{i \alpha \phi(z,\bar z)}$ and define the state
$\ket{\alpha}\equiv \mathcal{V}_\alpha(0,0)\ket{0}$. The Hilbert space can be constructed by application $a_{-n}$,$\bar a_{-n}$ (n $>$ 0) on such states.\\
The energy-momentum tensor is related to the generators of the conformal transformation $Q_\epsilon$ on the complex plane by
\begin{eqnarray}
Q_\epsilon=\frac{1}{2\pi i} \oint dz \epsilon(z) T(z),
\end{eqnarray}
where $\epsilon$ is an arbitrary holomorphic function. For any operator $\Phi(z)$
\begin{eqnarray}
\delta_\epsilon \Phi(z)=-[Q_\epsilon, \Phi(z)].
\end{eqnarray}
The energy-momentum tensor can be expanded as
\begin{eqnarray}
T(z)=\sum_{n\in \mathbb{Z}}z^{-n-2}L_n,
\end{eqnarray}
where $L_n$ are the local conformal transformations on the Hilbert space, which satisfies the celebrated Virasoro algebra,
\begin{eqnarray}
[L_n,L_m]=(n-m)L_{n+m}+\frac{c}{12}n(n^2-1)\delta_{n+m,0},
\end{eqnarray}
where $c$ is the central charge of the theory. For the anti-holomorphic part we could define similar generators $\bar L_n$, which commutates with the holomorphic ones, i.e., $[L_n,\bar L_m]=0$. The global conformal transformation corresponds to a subalgebra $sl(2,\mathbb{C})$  of Virasoro algebra, which contain $\{L_1,L_{-1},L_{0}\}$. The Hilbert space by application of $L_{-n}, \bar L_{-m}$($n\ge 1$ $m\ge 1$ ) forms a representation of Virasoro algebra, which is often called a Verma module $\mathcal{V}_{h,\bar h}$. The state $\ket{h,\bar h}$ is the eigenstate of generators $L_0, \bar L_0$, i.e., $L_0\ket{h,\bar h}=h\ket{h,\bar h}$ and $\bar L_0\ket{h,\bar h}=\bar h\ket{h,\bar h}$.
\subsection{Group coherent state and local excitation}
Coherent states defined for compact and non-compact Lie group have been well studied by many literatures\cite{Perelomov}\cite{Gilmore}\cite{Perelomov2}. The coherent states for quantum harmonic oscillator is one example. To construct the group coherent state we should have a Hilbert space, and the representation of the group
$G$ on it, which is denoted by $T(g)$ for any $g\in G$. As the quantum harmonic oscillator we should choose one reference vector $\ket{\psi_0}$, e.g., $\ket{\psi_0}$ in the quantum harmonic oscillator, in Hilbert space. Then choose the subgroup $G_0$ which satisfies
\begin{eqnarray}
T(h)\ket{\psi_0}=e^{i\phi(h)}\ket{\psi_0},
\end{eqnarray}
where $h\in G_0$ and $\phi(h)$ is the phase. The coherent state is defined by the element of coset space $G/G_0$, for $z(g)\in G/G_0$ we have
\begin{eqnarray}
\ket{x(g)}=T(x(g))\ket{\psi_0}.
\end{eqnarray}
What we will use in the following is the non-compact group $SL(2,\mathbb{C})/\mathbb{Z}_2$, for the element $g$
\begin{eqnarray}
g: z\to \frac{az+b}{cz+d}, \ \ \text{with} \ ad-bc=1.
\end{eqnarray}
$g$ is associated with a matrix
\begin{eqnarray}\label{Fundmentalrepresentation}
g=\left(\begin{array}{cc}
a\  b\\
c\  d
\end{array}\right).
\end{eqnarray}
The corresponding algebra $sl(2,\mathbb{C})$ satisfies the communication relation,
\begin{eqnarray}
[L_{1},L_{-1}]=2L_0,\ \ [L_{\pm 1},L_0]=\pm L_{\pm 1},\quad [\bar L_{1},\bar L_{-1}]=2\bar L_0,\ \ [\bar L_{\pm 1},\bar L_0]=\pm \bar L_{\pm 1}.
\end{eqnarray}
We review the fundamental representation of $sl(2,\mathbb{C})$ in the appendix. The result is
\begin{eqnarray}
L_0+\bar L_{0}=\Sigma_3,\quad L_1+\bar L_1=\Sigma_1+\sigma_2,\quad L_{-1}+\bar L_{-1}=\sigma_2-\Sigma_1,
\end{eqnarray}
\begin{eqnarray}
i(L_0-\bar L_0)=\sigma_3,\quad i(L_1-\bar L_1)=\Sigma_2-\sigma_1,\quad i(L_{-1}-\bar L_{-1})=\Sigma_2+\sigma_1,
\end{eqnarray}
where the matrices $\sigma_i$ and $\Sigma_j$ are defined as (\ref{FundamentalSL2c}).
The matrix (\ref{Fundmentalrepresentation}) can be parameterized as
\begin{eqnarray}
g= \left(\begin{array}{cc}
1\  0\\
z\  1
\end{array}\right)
\left(\begin{array}{cc}
1\  u\\
0\  1
\end{array}\right)\left(\begin{array}{cc}
e^{v}\  \ 0\\
0\  \ \ e^{-v}
\end{array}\right),
\end{eqnarray}
which can be expressed by the basis (\ref{FundamentalSL2c}) as
\begin{eqnarray}
g=e^{x(\Sigma_1-\sigma_2)+y(\Sigma_2+\sigma_1)}e^{u_1(-\Sigma_1-\sigma_2)+u_2(\Sigma_2-\sigma_1)}e^{v_1\Sigma_3+v_2\sigma_3},
\end{eqnarray}
where $x,y,u_1,u_2,v_2,v_2$ are real, $z\equiv x+iy, u\equiv -u_1-iu_2 $ and $v\equiv -\frac{1}{2}(v_1+iv_2)$. The general group element can be written as
\begin{eqnarray}
g=e^{zL_{-1}+\bar z \bar L_{-1}}e^{uL_1+\bar u \bar L_1}e^{vL_0 +\bar vL_0},
\end{eqnarray}
where $\bar z,\bar u, \bar v$ are the complex conjugation of $z,u, v$ respectively.
 The Verma module $\mathcal{V}_{h,\bar h}$ is a representation of $SL(2,\mathbb{C})$ group. Only part of the states in a
Verma module constitute the irreducible representation. These state are the ones obtained by application of the operator $L_{-1}$($\bar L_{-1}$) on
the state $\ket{h,\bar h}$, but not $L_{-n}$($n\ge 2$). To construct the coherent state we need to choose a reference state in the representation space.
Here the most natural one is the state $\ket{h,\bar h}$.\\

Now we turn to another different topic, the local excitation in 2D CFT, and see how to relate it to the group coherent state.
As we have mentioned in section \ref{setup} the local excitation of a primary operator $O$ is given by $O(\omega,\bar \omega)\ket{0}$. If assuming the spcacetime
is cylinder and making the map (\ref{radialmap}), we obtain a state on the complex plane,
\begin{eqnarray}\label{localexictationforgeneral}
\ket{z,\bar z}\equiv O(z,\bar z)\ket{0}.
\end{eqnarray}
The generator of the global conformal transformation $L_{-1}$ has the following commutation relation with the primary operator
\begin{eqnarray}
[L_{-1},O(z,\bar z)]=\frac{d}{dz}O(z,\bar z),
\end{eqnarray}
also for the anti-holomorphic part. So $L_{-1}$ and $\bar L_{-1}$ is the generator of translation on the complex plane. We have
\begin{eqnarray}\label{Coherentstateforgeneralcase}
O(z,\bar z)=e^{zL_{-1}}e^{\bar z\bar L_{-1}} O(0,0)e^{-zL_{-1}}e^{-\bar z\bar L_{-1}}.
\end{eqnarray}
Thus the local excitation can be rewritten as
\begin{eqnarray}\label{localexcitation}
\ket{z,\bar z}=e^{zL_{-1}}e^{\bar z\bar L_{-1}} O(0,0)\ket{0}=e^{zL_{-1}}e^{\bar z\bar L_{-1}}\ket{h,\bar h},
\end{eqnarray}
where we have used $L_{-1}(\bar L_{-1})\ket{0}=0$.
For the representation of the $SL(2,\mathbb{C})$ on the Verma module, if we choose the reference state to be $\ket{h}$,
the subgroup $H$ whose element $g_0$ satisfies
\begin{eqnarray}
g_0\ket{h,\bar h}=e^{i\phi(g_0)}\ket{h,\bar h},
\end{eqnarray}
is
\begin{eqnarray}
g_0=e^{uL_1+\bar u \bar L_1}e^{vL_0 +\bar vL_0}.
\end{eqnarray}
The element $x(g)$ of the coset space  $M_h=SL(2,\mathbb{C})/H$ is
\begin{eqnarray}
x(g)=e^{zL_{-1}+\bar z \bar L_{-1}}.
\end{eqnarray}
Therefore the coherent state for $SL(2,\mathbb{C})$ in the Verma module is defined as
\begin{eqnarray}\label{CoherentSL@C}
x(g)\ket{h,\bar h}=e^{zL_{-1}+\bar z \bar L_{-1}}\ket{h,\bar h},
\end{eqnarray}
 which is just the expression (\ref{localexcitation}). The local excitation of primary operator (\ref{localexcitation}) can be explained as the group coherent state for $M_h\otimes M_{\bar h}$
on the Verma module. The parameters of the coherent state is related to the coordinate of the local excitation.  \\
Unlike the coherent state for
the quantum harmonic oscillator the coherent state (\ref{CoherentSL@C}) is not the eigenvector of the generator $L_{+1}$ or $L_0$. On the other hand
one could also use the definition of the coherent state as the eigenvector of generator $L_{+1}$, i.e., $L_{+1}\ket{z}=z\ket{z}$. This state
exists, but it is not related to the local excitation of the primary operator.   \\
The recent paper discuss the local excitation of descendant operator\cite{Chen:2015usa}\cite{Caputa:2015tua}. We could obtain such coherent state by choosing the reference state smartly.

\section{Entanglement property of the coherent state}\label{EEForRCFT}
In this section we would like to study the entanglement between the left and right side of the local excitation position at  $t=0$. In the
free scalar case the defined entangled coherent state (\ref{Bellcoherentstatetime}) or (\ref{Deformedentangledcoherentstate}) do contribute
entanglement, which is related to the entanglement between the left and right moving modes. For the general rational CFT this is also
confirmed in paper \cite{He:2014mwa} by calculating the entanglement and R\'enyi entropy. Physically we could explain the phenomenon as the local excitation
produces two quasi-particle, which move left and right respectively. The first thing we want to show is that the entanglement is only related to the state $\ket{h,\bar h}$. It seems that these two parts are independent
with each others. But we can't write $\ket{h,\bar h}$ as a product state like $\ket{h}\ket{\bar h}$ for general case. Otherwise the coherent state
(\ref{Coherentstateforgeneralcase}) or the local excitation (\ref{localexictationforgeneral}) will also be a product state, thus the entanglement
entropy should be zero. One of the exception is the local excitation $e^{i\phi(x,t)}\ket{0}$, for which we could write $\ket{h,\bar h}\equiv \lim_{z,\bar z \to 0}e^{i\phi(z,\bar z)}\ket{0}$ as $e^{i\phi(0)}\ket{0}e^{i\bar \phi(0)}\ket{0}$. The entanglement entropy is zero as expected. For the general
primary operator $O(z,\bar z)$ we can't write it as $O(z)\bar O(\bar z)$.\\
In the free theory the operators $O_1=e^{i\phi(z,\bar z)}$ and $O_2=e^{i\phi(z,\bar z)}+e^{-i\phi(z,\bar z)}$ have the same conformal dimension. Thus we
can define two states \begin{eqnarray}
\ket{h,\bar h;O_1}\equiv \lim_{z,\bar z\to 0}O_1(z,\bar z)\ket{0}, \\
\ket{h,\bar h;O_2}\equiv \lim_{z,\bar z \to 0} O_2(z,\bar z)\ket{0}.
\end{eqnarray}
The entanglement between holomorphic and anti-holomorphic part of state $\ket{h,\bar h;O_2}$ is not zero, but $\log 2$. This entanglement should contain
the information of the operator. We would like to show that the entanglement of the highest-weight state $\ket{h,\bar h}$ and the coherent state (\ref{localexcitation})are same. We assume the state $\ket{h,\bar h}$ is normalized. Schmidt decomposition implies that the state $\ket{h,\bar h}$ can be expressed as
\begin{eqnarray}
\ket{h,\bar h}=\sum_{i,\bar j}\rho_{i\bar j}\ket{i}\ket{\bar j},
\end{eqnarray}
where $\ket{i}$ and $\ket{\bar j}$ are orthogonal bases for holomorphic and anti-holomorphic states respectively, and they satisfy
\begin{eqnarray} \label{Comunicationofschmit}
\bra{i}i'\rangle =\delta_{ii'},\ \ \bra{\bar j}\bar j'\rangle =\delta_{\bar j\bar j'},\ \ \bra{i}\bar j\rangle=0.
\end{eqnarray}
Since $L_0(\bar L_0) \ket{h,\bar h}=h(\bar h)\ket{h,\bar h}$, and $L_1(\bar L_1)\ket{h,\bar h}=0$, we also expect $L_0(\bar L_0)\ket{i}=h(\bar h)\ket{i}$,
and $L_1(\bar L_1)\ket{\bar j}=0$. After application of $e^{zL_{-1}}e^{\bar z \bar L_{-1}}$ on $\ket{h,\bar h}$, the relation (\ref{Comunicationofschmit}) keeps the same. One could check the fact by using the formula $e^{L_{+1}}e^{L_{-1}}=e^{L_{-1}}e^{-L_{+1}}e^{2L_0}$. So
the coherent state has the same entanglement entropy as the reference state.\\
The state $\ket{h,\bar h}$, as the highest representation state in the Verma module, gives us nothing about the entanglement property
of the holomorphic and anti-holomorphic part.  We will argue the entanglement property of the state $O(z,\bar z)\ket{0}$ is related to the structure of the  correlation function of
the operator  $O(z,\bar z)$. The higher-order correlation function  is in general written as a sum of product of holomorphic and anti-holomorphic functions. For the free scalar theory, the higher-order correlation function of operator $O_1=e^{i\phi}$ is just a product of holomorphic and anti-holomorphic function, which is different from $O_2=e^{i\phi}+e^{-i\phi}$. On the other hand we know the state $O_1\ket{0}$ is not entangled, which
is different from the state $O_2\ket{0}$. To see the relation of the two fact we need to know the OPE property of the operators, which leads to the
structure of higher-order correlation function.

\subsection{OPE and local excitation}
The local excitation is obtained by inserting a primary operator at point $x$ on a time slice. One could also insert another operator at the nearby point  on same time, thus get a new state, which can be formally written as $O_a(y)O_b(x)\ket{0}$. More generally we have the state  $\ket{\Phi_n}=O_1(x_1)O_2(x_2)...O_n(x_n)\ket{0}$.  As we have noted the local excitation $O_a(x)\ket{0}$ can be
seen as two quasi-particles which propagate on converse direction. In the same way the state $O_a(x')O_b(x)\ket{0}$ can be seen as two pairs quasi-particles.  What we are interested in is the entanglement property between the quasi-particle pair. Since the Hilbert space of the quasi-particles are increasing as
we inserting a new operator, the entanglement entropy $S(\ket{\Phi_n}$ is also expected to be not decreasing, i.e., $S(\ket{\Phi_n}) \ge S(\ket{\Phi_{n-1}})$. \\
 The entanglement entropy is also expected to be a function of the operator, i.e., $S(\ket{\Phi_n})=F[f(O_1),\\ f(O_2),...,f(O_n)]$, where
  $f: O\to \mathbb{C}$, $O$ is an operator.  And it should also satisfy
 the additivity, which means  $F[f(O_1 O_2...O_n)]=F[f(O_1)]+F[f(O_2)]+...F[f(O_n)]$. Physically this means the quasi-particles pairs are independent
 with each others. The state $\ket{\Phi_n}$ is direct sum of $n$ pairs quasi-particles states.\\
 The product of two operators in conformal field theory admits fusion algebra,
 \begin{eqnarray}\label{OPE}
 O_a(z,\bar z) O_b(\omega,\bar \omega)= \sum_{c}g_{ab}^c (z-\omega)^{h_c-h_a-h_b}(\bar z-\bar\omega)^{\bar h_c-\bar h_a-\bar h_b} O_c(\omega,\bar \omega)+...,
 \end{eqnarray}
where
$g_{ab}^c$ is the three point correlation function coupling constant, $...$ denotes the descendants. In the same way one could write down the
 OPE of n operators. From the Hilbert space point of view, (\ref{OPE})
means the equality of the states $LHS\ket{0}$  and $RHS\ket{0}$. Thus the local excitation $\ket{\Phi_n}$ is related to summation of other states.
We could describe the OPE more elegantly by abstract notation of fusion rules, i.e.,
\begin{eqnarray}\label{twofusion}
O_a\times O_b =\sum_c\mathcal{N}_{ab}^c O_c,
\end{eqnarray}
where
\begin{eqnarray}
\mathcal{N}_{ab}^c=
\begin{cases}
0, &\text{iff}\ \  g_{ab}^c=0  \cr 1, &\text{others}\end{cases}.
\end{eqnarray}
Note that the fusion numbers $\mathcal{N}$ factorizes as $\mathcal{N}_{left}\times \mathcal{N}_{right}$. When we fuse several operators we are
free to choose the order of the fusion. This is a consequence of the associativity of the OPE of primary fields. The space of states that
corresponds to the fusion process can be taken as a Hilbert space. The order of the fusion can be seen as the choice of the basis of such
Hilbert space. Different basis can be related to each other by the fusion matrix. All these are just the property that should be  satisfied by
the anyons, see, e.g., \cite{Pachos}\cite{Preskill}. In some way we can take the local excitation as creation of anyons, different operators correspond to different
charge of the anyons. Two anyons can fuse to other charges anyons, the fusion algebra behaves as the selection rule of the fusion process. We
will use the property of anyons to obtain the entanglement entropy of the local excitation.\\
Assume a physical process in which a pair $a\bar a$ are created, where $a$ labels the anyon, $\bar a$ is its antiparticle. The anyon could be self-conjugation,
$a=\bar a$. The quantum dimension $d_a$ is an important concept of the anyons $a$. $1/d_a$ is the amplitude for the annihilation of $a$ and its antiparticle $\bar a$. Thus $1/d_a^2$ is the probability that annihilation occurs, denoted as $p(a\bar a\to 1)=1/d_a^2$\cite{Preskill}. This property could be generalized to fusion process
of two anyons, $a$ and $b$. If the fusion rule is
\begin{eqnarray}
a\times b=\mathcal{N}_{ab}^c c,
\end{eqnarray}
the probability of the fusion $ab\to c$ is
\begin{eqnarray}\label{fusionpro}
p(ab\to c)=\frac{\mathcal{N}_{ab}^cd_c}{d_ad_b}.
\end{eqnarray}
Since the quantum dimension has the equality
\begin{eqnarray}
\sum_c \mathcal{N}_{ab}^c d_c =d_a d_b,
\end{eqnarray}
the probability $p(ab\to c)$ satisfies the normalization condition. For the local excitation operators we only consider the class in which $O_a\times O_a =1+...$, so the anyon produced by $O_a$ is self-conjugation.\\

\subsection{Entanglement entropy of local excitation}\label{EEforgeneraloperator}
We consider the rational CFT and assume inserting $n$ same operators $O_a$ at some region and obtain the state $\ket{\Phi_n}$. The entanglement entropy between the left and right chiral sector of such state $S(\ket{\Phi_n})=n S(\ket{O_a})$, where $\ket{O_a}$ is defined by $O_{a}\ket{0}$. By the OPE argument
\begin{eqnarray}\label{nfusion}
O_a\times O_a...\times O_a=\sum_{b_1,b_2,...,b_{n-2},c}\mathcal{N}_{aa}^{b_1}...\mathcal{N}_{b_{n-2}a}^{c}O_c,
\end{eqnarray}
$\ket{\Phi_n}$ is also related to the superposition of some fusion states $O_c\ket{0}$. It is not easy to write down the OPE of the product $O_a....O_a$
and get the exact states. Here we just assume the result state is a superposition of the normalized state $\ket{c}\equiv N_{c} O_c \ket{0}$ with the weight $\lambda_c$, where $N_c$ is the normalization constant, and $\sum_{c}\lambda_c^2=1$,
 \begin{eqnarray}\label{1}
 \ket{\Phi_n}=\sum_c \lambda_c \ket{c}.
 \end{eqnarray}
 The states $\ket{c}$  are orthogonal to each other. $\ket{\Phi_n}$ can be
explained as the state for $n$ anyons with charge $a$. These anyons fuse to other charges $c$, but the probability is not same for different charges(\ref{fusionpro}). This process is shown in figure 1.\\

\begin{figure}[htbp]
\centering\includegraphics[width=2.5in]{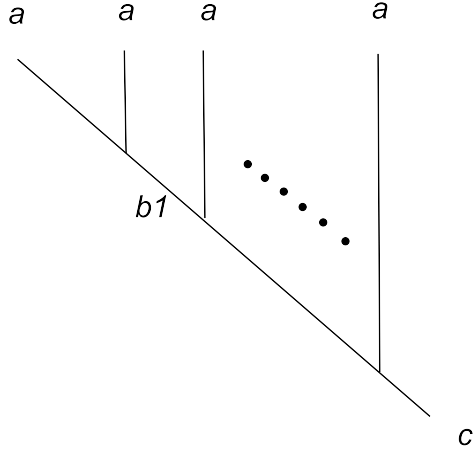}
\caption{The fusion process}\label{fig:1}
\end{figure}

One could divide the process into $n-1$ times fusion, which are labeled as states
\begin{eqnarray}
\ket{aa\to b_1}\ket{a b_1\to b_2}...\ket{a b_{n-2}\to c}.
\end{eqnarray}
These states constitute as the basis of the Hilbert space of the fusion process.\\
The probability of each fusion result is given by (\ref{fusionpro}), each step is independent with others. Thus the probability that $\ket{\Phi_n}$ fuses
to a fixed $\ket{c}$ is
\begin{eqnarray}
p(aa...a\to c)=\sum_{b_1,b_2,b_{n-2}}\frac{\mathcal{N}_{aa}^{b_1}d_{b_1}}{d_a^2}\times \frac{\mathcal{N}_{ab_1}^{b_2}d_{b_2}}{d_a d_{b_1}}\times ...\times \frac{\mathcal{N}_{ab_{n-2}}^cd_c}{d_ad_{b_{n-2}}}=\sum_{b_1,...,b_{n-2}}\frac{\mathcal{N}_{aa}^{b_1}...\mathcal{N}_{ab_{n-2}}^cd_c}{d_a^n}.
\end{eqnarray}
One could see from (\ref{1}) $\lambda_c^2=p(aa...a\to c)$. Now it is ready to get the entanglement entropy of the RHS  of (\ref{1}). Let's denote the
entanglement entropy of the state $\ket{c}$ as $S(\ket{O_c})$, the result is
\begin{eqnarray}\label{S}
S=-\sum_c\lambda_c^2\log \lambda_c^2+\sum_{c}\lambda_c^2 S(\ket{O_c}).
\end{eqnarray}
We argue that the entanglement entropy of LHS of (\ref{1}) is $nS(\ket{O_a})$, which is only depended on $O_a$. So $S$ can't depend on $O_c$. Let's
rewrite (\ref{S}) as
\begin{eqnarray}\label{key}
S=nS(\ket{O_a})=\sum_{c}\sum_{b_1,...,b_{n-2}}\frac{\mathcal{N}_{aa}^{b_1}...\mathcal{N}_{ab_{n-2}}^cd_c}{d_a^n} (S(\ket{O_c})-\log \frac{\mathcal{N}_{aa}^{b_1}...\mathcal{N}_{ab_{n-2}}^cd_c}{d_a^n} ).
\end{eqnarray}
For rational CFT $\mathcal{N}=1$ if the fusion process could happen. Note that for the fusion process (\ref{nfusion}), one has
\begin{eqnarray}
d_a^n=\sum_{b_1,...,b_{n-2},c}\mathcal{N}_{aa}^{b_1}...\mathcal{N}_{ab_{n-2}}^cd_c.
\end{eqnarray}
Thus the solution of (\ref{key}) is
\begin{eqnarray}\label{Solution}
 S(\ket{O_c})=\log d_c,\quad S(\ket{O_a})=\log d_a.
 \end{eqnarray}
 This is also consistent with our previous conclusion that
the entanglement entropy of the local excitation should only depend on the information of the operator.\\
Let's comment on the proof before we turn to an example, the Ising model. Since the local excitation is usually ill defined one has to use some method
to regularize such state. So the n product state $\ket{\Phi_n}$ also needs some regularization. In principle by suitable regularization one could
write down the OPE of the n product operators, i.e., RHS of (\ref{nfusion}). Thus the weight $\lambda_c$ of the state $\ket{c}$ could also be obtained, which should be related to the OPE coefficients. Here we just assume the local excitation behaves as the anyons, and use the fusion probability. We
expect it would help to find a suitable regularization method to define the n product state $\ket{\Phi_n}$.\\
The solution (\ref{Solution}) relies on the fact $\mathcal{N}=0$ or $1$ for rational CFT. For some theories this will not be true. It is not easy to
find such simple solution (\ref{Solution}).
\subsection{Ising model}
The critical Ising model in 2D can be described by the simplest nontrivial unitary minimal model, $\mathcal{M}(4,3)$ .
There are three primary operators in Ising model: the spin $\sigma$, the energy density $\epsilon$ and the identity $I$.
The quantum dimension of $\sigma$ is $\sqrt{2}$, and $1$ for $I$, $\epsilon$. The local excitation of $\sigma$ is a
nontrivial example to check the result in sect. (\ref{EEforgeneraloperator}). In this section we would like to use
another view to get the entanglement entropy for local excitation of $\sigma$. The Ising model can also be described by
a free massless real fermion. In term of the Fermion field the spin operator is nonlocal. But one could start from two decoupled Ising model, which are
equal to free massless scalar theory(\ref{freetheorymassless}).  The product of the spin operators $\sigma_1(x)\sigma_2(x)$ is related to the free massless scalar theory as
\begin{eqnarray}
\sigma_1(z,\bar z)\sigma_2(z,\bar z)=\sin \phi(z,\bar z),
\end{eqnarray}
where $\sigma_1$, $\sigma_2$ are spin operators of the two decoupled Ising model, $\phi(z,\bar z)$ is the free scalar field. The entanglement entropy of
the operator $\sin \phi\equiv (e^{i\phi}-e^{-i\phi})/2i$ is $\log 2$ as we have shown in sect.(\ref{Freeentangledcoherentstate}). The entanglement
of the state $\sigma_1 \sigma_2\ket{0}$ is equal to $2S(\sigma_1\ket{0})$. Thus the entanglement of the state $\sigma_1\ket{0}$ is $\log \sqrt{2}$, this is consistent with our general result (\ref{GR}). Similarly, the product of the  energy operator can be expressed as
\begin{eqnarray}
\epsilon_1(z,\bar z)\epsilon_2(z,\bar z)=4\partial \phi \bar \partial \phi.
\end{eqnarray}
The entanglement entropy of the state $\epsilon_1\ket{0}$ is $0$, this is also consistent with the fact $d_\epsilon=1$.

\section{Comment on the holographic view}\label{Comment}
The radial quantization can be used for any dimension CFT \cite{Pappadopulo:2012jk}. Thus one could construct the correspondence between the local
excitation and coherent state in any dimension. Here we still focus on the 2D CFT, with the dual 3D gravity theory.  Recently many authors study the
correspondence in CFT of  local excitation state of the field theory in bulk. Our question here is actually converse to theirs, i.e., what is the
correspondence in the bulk of the local excitation in the boundary CFT.  The coherent state behaves as a the most classical
state as we have mentioned\cite{Zhang}\cite{Hepp}\cite{Yaffe}. In $AdS/CFT$ the CFT has a classical gravity dual, in the leading $N$ order, the physical
qualities are expected to be related to the geometry in the bulk spacetime. The bulk and boundary have the same symmetry group, e.g., in the 2D, the global conformal transformation $SL(2,\mathcal{C})$. \\
The primary state $\ket{h,\bar h}\equiv \lim_{z\to 0\bar z\to 0}O(z,\bar z)\ket{0}$ has the energy $E\propto h+\bar h$. It should be an eigenvector of the
dilation generator $D$. In the bulk the dilation operator is actually the Hamiltonian. The translation operator $P_\mu$ and special conformal transformation (SCT) operator $K_\mu$ behaves as the raising and lowering operator of the energy. One could solve the Schr\"odinger equation of the particle of mass $m$ in AdS spacetime. The lowest energy state $\ket{\psi_0}$ should satisfy the constraint $K_{\mu }\ket{\psi_0}=0$, the same constraint also appears
in Hilbert space related to primary state $\ket{h,\bar h}$, $L_{+1}(\bar L_{+1})\ket{h,\bar h}=0$. The object at rest in $AdS$ with suitable mass $m$ is expected to correspond to the primary state $\ket{h,\bar h}$ in boundary $CFT$\cite{Kaplan}. The local excitation in CFT at any point can be constructed
as the coherent state of the group $SL(2,\mathcal{C})/ Z_2$ choosing the reference state $\ket{h,\bar h}$. The corresponding bulk state is just the similar
coherent state constructed by the bulk symmetry group choosing the reference state as the stationary particle. Thus the time evolution of the
local excitation in  CFT is related to the particle motion in AdS. The model that constructed in paper \cite{Nozaki:2013wia} realizes the property we
expected. But it is still interesting to investigate the detail of the correspondence, such as the relation between  mass $m$ and the conformal dimension $h$. We will work more on this in the future.
\section{Conclusion}
In this paper we investigate the coherent state and local excitation in 2D CFT. We find the relation between the local excitation and a group coherent state. For rational 2D CFT we obtain the entanglement entropy between the left and right sector for local excitation of a primary operator $O_a$, which is related to the quantum dimension of this operator. The result is consistent with the calculation by using the replica trick. Our approach is based on
the assumption that the local excitation behaves as the anyon. This is reasonable since the operators have the similar fusion relation like the anyons.  This approach is simple and more physical. In principle one also could use this method to find the result for theories beyond rational CFT. In some
sense our method also confirm the similarity between the local excitation and anyon. It may be interesting to use the local excitation to imitate the
anyon in field theory. Many topics related to anyons are worthy to investigate in such system. \\
The relation we find between the local excitation and a group coherent state is more than a mathematical equality.
When considering the gravity dual of local excitation, we believe it provide a new view on this. Since the symmetry of the bulk and boundary CFT is same,
our definition of the local excitation by a group coherent state can be translated to bulk directly. The coherent state usually behaves as the ``most
classical'' state  in a quantum theory. The AdS description of the boundary CFT actually is a classical system in the leading $N$ order.
It is interesting to investigate whether a properly defined coherent state in the boundary CFT should have a correspondence in its classical gravity
dual. \\
Besides these we also study the so-called Glauber coherent state in 2D free boson theory. We study the property of the coherent state, and the
entangled coherent state, which evolutes under the free or time-dependent Hamiltonian (\ref{firstorderH}). We also define the deformed local excitation, which is
generalized concept of local excitation. Using this state we study the entanglement entropy of (\ref{DeformedCoherentEntangled}), we also show that
the violation of Bell inequality of such state. For the general CFT one can't explain the entanglement as bipartite system, since the maximal entanglement entropy of bipartite system is $\log 2$. In many theories
one could find operator with quantum dimension $d_a\ge 2$. It is a challenge to prove the violation of Bell's inequality for general case, since it is expected to be n-partite entanglement. \\
The concept of the coherent state in field theory is not new, but the entanglement property of the coherent state, and the possible gravity dual of
the coherent state is quite interesting topic. Recently the entanglement property of boundary state in CFT is investigated in many papers, 
such as \cite{PandoZayas:2014wsa}-\cite{Miyaji:2015fia}. At least for the free scalar or fermion the boundary state is one kind of coherent state\cite{PandoZayas:2014wsa}. For the rational CFT the universal 
term of the left and right entanglement is also related to the quantum dimension\cite{Das:2015oha}. It is also shown in \cite{Miyaji:2014mca}-\cite{Miyaji:2015fia} one kind of the boundary state 
corresponds to the trivial spacetime. It seems there are some relation between such different concepts. The coherent state may be a key point to understand this.   

\vskip 0.5cm
{\bf Acknowledgement}
\vskip 0.2cm
We are grateful to Song He, Miao Li, Masamichi Miyaji, Tokiro Numasawa, Tadashi Takayanagi and  Kento Watanabe for useful conversations and discussion. Specially we would like to thank Tadashi Takayanagi for pointing out some mistakes.  We thank Miao Li, Tadashi Takayanagi for their encouragement and support.  The author is supported by Postgraduate Scholarship Program of China Scholarship Council.

\appendix
\section{Group $SL(2, \mathbb{C})$}
In this appendix we review some properties of group $SL(2,\mathbb{C})$ and its fundamental representation. In the fundamental representation $SL(2,\mathbb{C})$ is the non-degenerate complex matrices with unit determinant. The
dimensions of the group is 6, and the Lie algebra consists of traceless complex matrices. The real group $SL(2,R)$, which has the dimension 3, is a subgroup of $SL(2,\mathbb{C})$. One could choose the following basis for
the Lie algebra of $SL(2,\mathbb{C})$ in the fundamental representation,
\begin{eqnarray}\label{FundamentalSL2c}
\sigma_1=\frac{1}{2}\left(\begin{matrix}0 & i\\ i&0 \end{matrix}\right),\quad \sigma_2=\frac{1}{2}\left(\begin{matrix}0 & 1\\ -1& 0 \end{matrix}\right),\quad \sigma_3=\frac{1}{2}\left(\begin{matrix}-i & 0\\ 0& i \end{matrix}\right), \\ \nonumber
\Sigma_1=\frac{1}{2}\left(\begin{matrix}0 & 1\\ 1& 0 \end{matrix}\right),\quad \Sigma_2=\frac{1}{2}\left(\begin{matrix}0 & -i\\ i& 0 \end{matrix}\right),\quad \Sigma_3=\frac{1}{2}\left(\begin{matrix}-1 & 0\\ 0& 1 \end{matrix}\right).
\end{eqnarray}
One could check the commutation relation
\begin{eqnarray}
[\sigma_i,\sigma_j]=\epsilon_{ijk}\sigma_k,\quad
[\sigma_i,\Sigma_j]=\epsilon_{ijk}\Sigma_k,\quad
[\Sigma_i,\Sigma_j]=-\epsilon_{ijk}\sigma_k.
\end{eqnarray}
The algebra associated with the global conformal group of 2D CFT is generated by $L_{-1},L_0, L_1$ and $\bar L_1,\bar L_0,\bar L_{-1}$, which satisfy the
commutation relation
\begin{eqnarray}\label{conformalgroupgenerator}
[L_i,L_j]=(i-j)L_{i+j}, \quad \ \ [\bar L_i,\bar L_{j}]=(i-j)\bar L_{i+j},\quad \ [L_i, \bar L_j]=0,
\end{eqnarray}
where $i,j=-1,0,1$.
In the fundamental representation the generators $L_i$ and$\bar L_j$ can be expressed by the basis (\ref{FundamentalSL2c}) as
\begin{eqnarray}
L_0+\bar L_{0}=\Sigma_3,\quad L_1+\bar L_1=-\Sigma_1-\sigma_2,\quad L_{-1}+\bar L_{-1}=\Sigma_1-\sigma_2,
\end{eqnarray}
\begin{eqnarray}
i(L_0-\bar L_0)=\sigma_3,\quad i(L_1-\bar L_1)=\Sigma_2-\sigma_1,\quad i(L_{-1}-\bar L_{-1})=\Sigma_2+\sigma_1.
\end{eqnarray}
One could check above relations by (\ref{FundamentalSL2c}) and (\ref{conformalgroupgenerator}).

\end{document}